\documentclass{agujournal2025}
\usepackage{url} 
\usepackage{lineno}
\usepackage[inline]{trackchanges} 
\usepackage{soul}
\usepackage{xcolor}
\usepackage{amsmath}
\usepackage[hidelinks]{hyperref}
\usepackage{orcidlink}
\usepackage{natbib} 
\usepackage{outlines}
\usepackage{datetime2}

\newcommand{\MA}{Mach number}

\draftfalse
\journalname{Geophysical Research Letters}

\begin{document}

\title{Averaging Effects on the Solar Wind Alfv\'en Mach Number: 
Implications for 
Switchbacks and Alfv\'en Transition}

\authors{
Joshua Goodwill      \affil{1}\orcidlink{0000-0002-5354-1164},
Subash Adhikari      \affil{1}\orcidlink{0000-0003-2160-7066},
Francesco Pecora     \affil{1}\orcidlink{0000-0003-4168-590X},
Rayta Pradata        \affil{1}\orcidlink{0009-0005-9366-6163},
Jiaming Wang         \affil{1}\orcidlink{0009-0008-8723-610X},
Sujan Prasad Gautam  \affil{1}\orcidlink{0000-0001-7379-4268},
Manuel E. Cuesta     \affil{2}\orcidlink{0000-0002-7341-2992},
Orlando  Romeo      \affil{3}\orcidlink{0000-0002-4559-2199}
David Ruffolo        \affil{4}\orcidlink{0000-0003-3414-9666},
Panisara Thepthong   \affil{5}\orcidlink{0000-0001-9597-1448},
Peera Pongkitiwanichakul \affil{6}\orcidlink{0000-0002-6609-1422},
Rohit Chhiber        \affil{1,7}\orcidlink{0000-0002-7174-6948},
William H. Matthaeus \affil{1}\orcidlink{0000-0001-7224-6024}
}

\affiliation{1}{Department of Physics and Astronomy, University of Delaware, Newark, DE 19716, USA}
\affiliation{2}{Department of Astrophysical Sciences, Princeton University, Princeton, NJ 08544, USA}
\affiliation{3}{Space Sciences Laboratory, University of California, Berkeley, CA 94720, USA}
\affiliation{4}{Department of Physics, Faculty of Science, Mahidol University, Bangkok 10400, Thailand}
\affiliation{5}{LPC2E, OSUC, CNRS, University of Orléans, CNES, Orléans F-45071, France}
\affiliation{6}{Department of Physics, Faculty of Science, Kasetsart University, Bangkok 10900, Thailand}
\affiliation{7}{Heliophysics Science Division, NASA Goddard Space Flight Center, Greenbelt, MD 20771, USA}

\correspondingauthor{Joshua Goodwill}{goodwill@udel.edu}
\keypoints%
{
Time-averaging interval should be chosen according to problem under consideration and can impact
characterization of sub-Alfv\'enic intervals
}
{
Switchbacks tend to be associated with the super-Alfv\'enic regime when Alfv\'en Mach number is computed with 
appropriate
averaging period
}
{Choosing long averaging periods can obscure brief excursions into a
super-Alfv\'enic state.
}

\begin{abstract}
Averaging techniques in solar wind measurements have been a longstanding subject of debate. Using Parker Solar Probe (PSP) observations from encounters 1–19, we investigate how averaging timescales influence the characterization of turbulent properties across the Alfv\'enic transition. We compute the rolling mean Alfv\'en Mach number over various averaging intervals, which are then analyzed against switchbacks, magnetic fluctuation energy, and correlation time. We find that the distribution of sub-Alfv\'enic intervals is relatively insensitive to judiciously-chosen averaging scales. In contrast, magnetic fluctuation energies increase systematically with larger averaging window, while maintaining a consistent profile across the Alfv\'en transition. We further show that the effective magnetic correlation time decreases with decreasing heliocentric distance and $M_A$, reaching values of several minutes approaching $M_A =1$. These results demonstrate the importance of choosing physically meaningful backgrounds for turbulence parameters, such as the correlation scales, and their impacts on characterizing the solar wind.
\end{abstract}

\section*{Plain Language Summary}
Parker Solar Probe (PSP) has recently measured the sub-Alfv\'enic zone, the region where the solar wind velocity is less than the propagation speed of fluctuations along magnetic fields.  This region has been shown to have different turbulence characteristics compared to other regions of the heliosphere. As PSP is measuring this new region of the solar wind, it is important to choose a suitable averaging length to properly characterize these measurements, especially with regard to evaluating the Alfv\'en Mach number ($M_A$).  In this study, we show how averaging can impact the Alfv\'en Mach number  as well as turbulent magnetic fluctuations. This study is particularly applicable to identifying switchbacks, or sudden magnetic reversals, and their abundance in the solar wind.

%
%

\section{Introduction}

The solar wind is a continuous stream of ionized plasma emitted from the Sun, whose properties evolve with heliocentric distance. One of the key parameters used to characterize its behavior is the Alfv\'en Mach number ($M_A$), which quantifies the ratio of bulk flow speed to a characteristic Alfv\'en speed. 
When the \MA\footnote{For brevity, hereafter 
for Alfv\'en \MA, we use the abbreviated nomenclature ``\MA'' where it does not cause confusion with regard to the sonic \MA.}
exceeds unity,
MHD disturbances propagate in the flow direction, i.e., outward, even for the backward propagating Alfv\'en wave. For this reason, super-Alfv\'enic solar wind is often associated with disconnection from the lower corona in the sense that Alfv\'enic
signals cannot propagate back to the source. 
The \MA\ plays a central role in the identification of shocks, helps characterize turbulence properties such as compressibility, turbulence amplitude, variance, switchbacks, and other features in collisionless plasmas \citep{kasper2019_Strong, bandyopadhyay2022_SubAlfvenic, zank2024_Characterization, goodwill2026_Parker, payne2026_Evolution}. 
The surface \citep[or zone, ][]{chhiber2022_extended, chhiber2024_Alfven}
at which the \MA\ exceeds unity is important in evaluating solar wind models such as WKB theory
\citep{ruffolo2024_Observed}.
Therefore, an accurate estimation of the \MA\ is essential for understanding the dynamical evolution of the solar wind. 

With the launch of Parker Solar Probe (PSP) \citep{fox2016_Solar}, an unprecedented amount of in situ measurements of the solar wind at heliocentric distances far closer to the sun have been made than previously possible \citep{raouafi2023_Parker}. 
Such observations make it possible to 
probe the solar wind properties
below and near the Alfv\'en transition.
Clearly, to properly diagnose this transition, it is necessary to systematically and accurately evaluate the \MA. The procedure for computing the \MA\ should be constructed keeping in mind the physical process under examination, and this consideration has led to some contrasting views of the conditions required for a 
proper determination
\citep{ruffolo2024_Observed, sioulas2025_Propagation}. This motivates the present study, which seeks to uncover the limitations of 
different choices of averaging intervals on determinations of \MA.
In particular, we attempt to clarify the influence of averaging on 
conditional statistics involving the 
\MA.

The PSP mission systematically explores the 
inner regions of solar wind in which the \MA\ changes from predominantly super-Alfv\'enic to predominantly sub-Alfv\'enic with decreasing heliocentric distance.
PSP is well-suited for this as its payload 
provides high-cadence measurements of velocity, density, temperature, and magnetic field strength. These observations allow for detailed computation of \MA s in regions where the solar wind transitions from sub-Alfv\'enic to super-Alfv\'enic flow \citep{bandyopadhyay2022_SubAlfvenic, zhao2022_Turbulence, zhao2022_Turbulent, adhikari2026_Characterization}.
While the overall picture of the 
Alfv\'enic transition is not controversial, 
characterization of specific time periods can sometimes be debated. 
At issue is the level of averaging and the sequence of averaging of the quantities that make up the \MA, i.e., density, 
magnetic field strength and flow speed. Discussion of this issue and how it affects the interpretation of the observations are the main points of the present paper. 

Because \MA\ calculations involve the flow speed, plasma density, and magnetic field strength, the influence of averaging can be nontrivial and potentially alter the characterization of certain plasma parcels. The averaging interval defines the background plasma state relative to which fluctuations are measured. Rolling (or moving) averages \citep{germano1992_Turbulence, bruno2013_Solar} are often applied to reduce instrumental noise, suppress high-frequency fluctuations, and extract large-scale trends. Such averaging, basically a low-pass filtering operation,  can improve the derived plasma parameters by smoothing sharp gradients and transient structures, making the distribution function of fluctuations more Gaussian in nature \citep{sorriso-valvo1999_Intermittency}. 
Ultimately, the choice of how $M_A$ is defined and evaluated, and in particular what sort of averaging is imposed, 
will depend on the context. For some applications, a more global averaging (bulk \MA) may be appropriate, for example, to characterize an entire region of the solar atmosphere, such as a particular solar wind stream \citep{sioulas2026_Generation}. In other contexts, it may be desirable to have a more local estimation (instantaneous \MA)
if the purpose is 
to characterize potential local dynamical processes, including, for example, instabilities \citep{chandrasekhar2013hydrodynamic}.
We revisit this in Section~\ref{sec:discussion}.

Although the computation of \MA\ 
appears to be elementary, there has been a surprising 
range of opinions regarding the question of 
what is the correct averaging interval to employ in this computation. 
Previous studies \citep{bandyopadhyay2022_SubAlfvenic, zhao2022_Turbulence, zhao2022_Turbulent, chhiber2024_Alfven, payne2026_Evolution, sioulas2026_Generation} characterizing the sub-Alfv\'enic wind have employed a broad range of averaging techniques and interval lengths. This variability raises the question of whether 
physical results remain consistent across different choices of averaging method. 

In this study, we investigate the sensitivity of \MA\  estimation due to varying the averaging intervals. 
In addition, we investigate the effect of rolling centered, boxcar averages on the magnetic field fluctuation energy to quantify its impact on non-WKB dynamics as previously analyzed by \citet{ruffolo2024_Observed}. 
Lastly, we assess the effective correlation time across the Alfv\'en transition to provide an estimation of a proper background mean in the turbulent solar wind. The results are intended to provide guidance for selecting appropriate windows for estimation of the \MA\  and other turbulent properties and to 
clarify
how this
choice influences
interpretation of solar wind dynamics at the Alfv\'en transition.

\section{Parker Solar Probe Encounter Data}\label{sec:PSP_data}
\subsection{Instruments}
The FIELDS instrument onboard PSP includes a flux-gate magnetometer (MAG) that measures the magnetic field ($\mathbf{B}$)~\citep{bale2016_FIELDS}. Electron density ($N_e$) is obtained from the quasi-thermal noise (QTN) spectrum captured by the Radio Frequency Spectrometer (RFS), which is part of the FIELDS suite~\citep{pulupa2017_Solar}. 
The Solar Wind Electrons Alphas and Protons (SWEAP) instrument suite~\citep{kasper2016_Solar} contains the Solar Probe Analyzers for Ions (SPAN-I), which measures 
the solar wind density ($N$) and velocity ({\bf V})
in the Radial–Tangential–Normal (RTN) coordinate system~\citep{livi2022_Solara}. 
We make use of the radial velocity component $V_R$, which is by far the dominant component, and is well measured by the instruments.
For this study, we use PSP data from encounters $E8-E19$ for all analysis except for Figure~\ref{fig:tau_c}, which uses encounters $E1-E19$. 

In order to calculate the Alfv\'en speed and Mach number, we determine the plasma density $N$ following the procedure of \citet{ruffolo2024_Observed}, 
which can be briefly summarized as follows:
When possible, we use a simplified heuristic approach by \citet{romeo2023_NearSun} to estimate $N$ (with 1-min cadence) as the QTN electron number density $N_e$, with the exception of E15, for which we use $N_e$ from the procedure of \citet{Moncuquet_2020}.
When $N_e$ is not available, and when the bulk of the solar wind proton distribution is within the field of view of SPAN-I, we instead use $0.86 n_p$, where $n_p$ is the proton density measured by SPAN-I and we multiply by an empirical factor of 0.86
\citep[derived from a comparison with $N_e$ at times when both measures are available;][]{ruffolo2024_Observed}.

For variables other than density, a $10\,\rm{s}$ cadence is used. 
To align the time cadence of datasets, $N_e$ was up-sampled to $10\,\rm{s}$ using linear interpolation. All figures, therefore, utilize a resolution of $10\,\rm{s}$ cadence.

\subsection{\MA, Switchback Parameter, and Fluctuation Energy}

Using the quantities from the instruments mentioned, we calculate the mean \MA\ ($M_A$) by
\begin{equation}
    \langle M_A \rangle_\tau= \frac{\langle V_R \rangle_\tau} {\langle V_A \rangle_\tau}
    \label{eq:MA}
\end{equation}
where $\langle \cdots \rangle_\tau$ denotes a rolling mean window average over an interval $\tau$, the mean Alfv\'en speed is $\langle V_A \rangle_\tau = \langle|\mathbf{B}|\rangle_\tau\left/ \sqrt{\mu_0 m_p \langle N_e \rangle_\tau}\right.$, $\mu_0$ is the magnetic permeability in a vacuum, $m_p$ is the proton mass, and $N_e$ is the electron number density. Results are largely unchanged when redefining the Alfv\'en speed as $\langle V_A \rangle_\tau = \langle |B|/\sqrt{\mu_0 m_p N_e} \rangle_\tau $.

The switchback parameter ($Z$) is a measure of magnetic deflection relative to the background field, defined as
\begin{equation}
    Z_\tau = \frac{1}{2} \left[ 1- \cos{ \left(\frac{\mathbf{B} \cdot \langle \mathbf{B}\rangle_{\tau}}{|\mathbf{B}| |\langle \mathbf{B}\rangle_{\tau}|} \right) }\right]
\end{equation}
where $\mathbf{B}$ is the  $10\,\rm{s}$ magnetic field, $\langle \mathbf{B} \rangle_{\tau}$ is the mean magnetic field and $Z \in [0,1]$. Here, $Z > 1/2$ indicates that the magnetic field is in a polarity-reversed state, and $Z < 1/2$ corresponds to ``background'' magnetic polarity. Throughout this paper, the term “switchback” refers exclusively to deflections that reverse the polarity, i.e., $Z > 1/2$. Note that for calculation of the switchback parameter $Z$, the ``background" magnetic field has previously been averaged over $\tau=6$ hours in ~\citet{dudokdewit2020_Switchbacks}, for a clear contrast between switchbacks and a long-duration average. For the present purposes, we vary $\tau$ over a range of $1\,\rm{min}$ to $10 \, \rm{hr}$, and find that a shorter background 
averaging interval tends to eliminate many switchbacks.  We therefore show $Z_{\rm 10hr}$ for the longest value $\tau=10$ hr in order to display the maximum number of switchbacks.

The mean-squared fluctuation energy for the magnetic field $\mathbf{B}$ is computed by
\begin{equation}
\delta B^2_\tau =  \sum_i \big( \langle B_i^2 \rangle - \langle B_i \rangle^2 \big)
\end{equation}
where $i$ refers to the RTN coordinate system. 

While currently we use rolling mean averaging for all data used in the present paper, it should be noted that we also compared these results using rolling median averages which results in minor differences, mainly pertaining to the magnitude of values but not their distributions.

\subsection{Correlation Time}
The normalized autocorrelation function for the magnetic field ($\mathbf{B}$) is defined by
\begin{equation}
    R(t') = \frac{\langle \mathbf{b}(t)\cdot \mathbf{b}(t+t')\rangle_{T-t'}}{\langle \mathbf{b}(t) \cdot \mathbf{b}(t) \rangle_T}
    \label{eq:auto}
\end{equation}
where $\mathbf{b} = \mathbf{B} - \langle \mathbf{B}\rangle_T$ is the magnetic fluctuation over a defined interval length $T$. 
The averaging
$\langle \cdots \rangle_{T - t'}$ in Eq.~\ref{eq:auto} refers to an average over the portion of the interval for which both times are contained within the interval $T$. \cite[see, e.g.,][]{matthaeus1982_Measurement,cuesta2022_Intermittency, roy2022_Karman}. 

The correlation time ($\tau_c$) is found by the ``e-folding" time ($R(\tau_c) = 1/e$), which is related to the size of the largest coherent structures. An interval length of $T=4\,\rm{hr}$ is used for consistency with previous studies \citep{parashar2020_Measures, cuesta2022_Isotropizationa} with a maximum lag ($t'_{max}$) of $90\,\rm{min}$ to allow for minimum required oversampling \citep{isaacs2015_Systematic, cuesta2022_Isotropizationa}. Correlation times have been calculated with varying interval length up to $6\,\rm{hr}$ and maximum lags up to $2\,\rm{hr}$, all maintaining similar values and trends.

\section{Results\label{sec:results}}

\begin{figure}
\includegraphics[width = \linewidth]{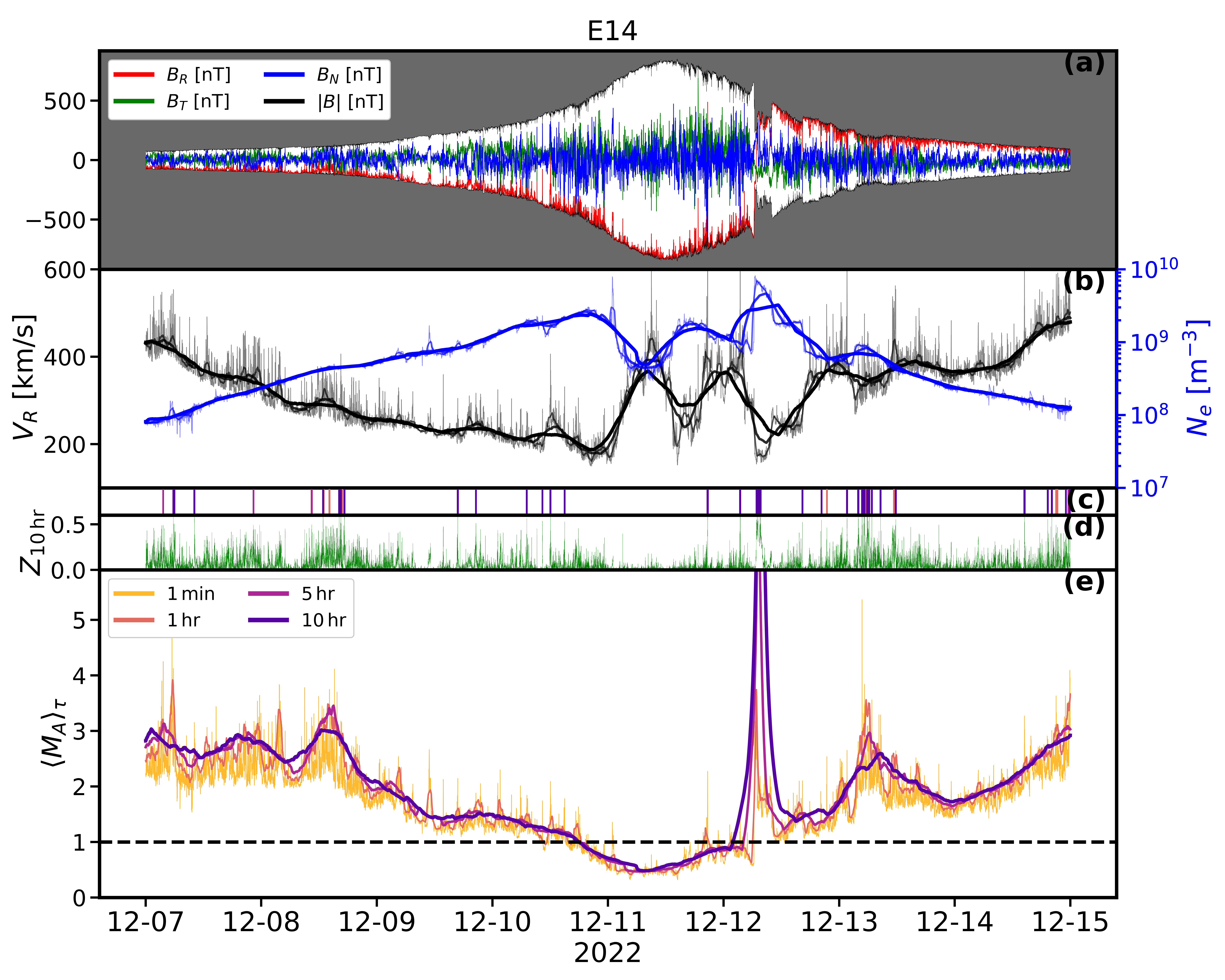}
\caption{
{
PSP $E14$ time series of (a) magnetic field components $(B_R, B_T, B_N)$ and magnitude $(|B|)$, (b) radial velocity $(V_R)$ and electron density $(N_e)$, (c) time in switchback state $(Z > 0.5)$, (d) switchback parameter relative to $10\,\rm{hr}$ average $(Z_{10 \rm{hr}})$, and (e) \MA\ $(M_A)$. Panels (b) and (e) display averages varying from $1\, \rm{min}$ to $10\, \rm{hr}$.
}
 }\label{fig:enc14}
\end{figure}

\begin{figure}
\includegraphics[width=\linewidth]{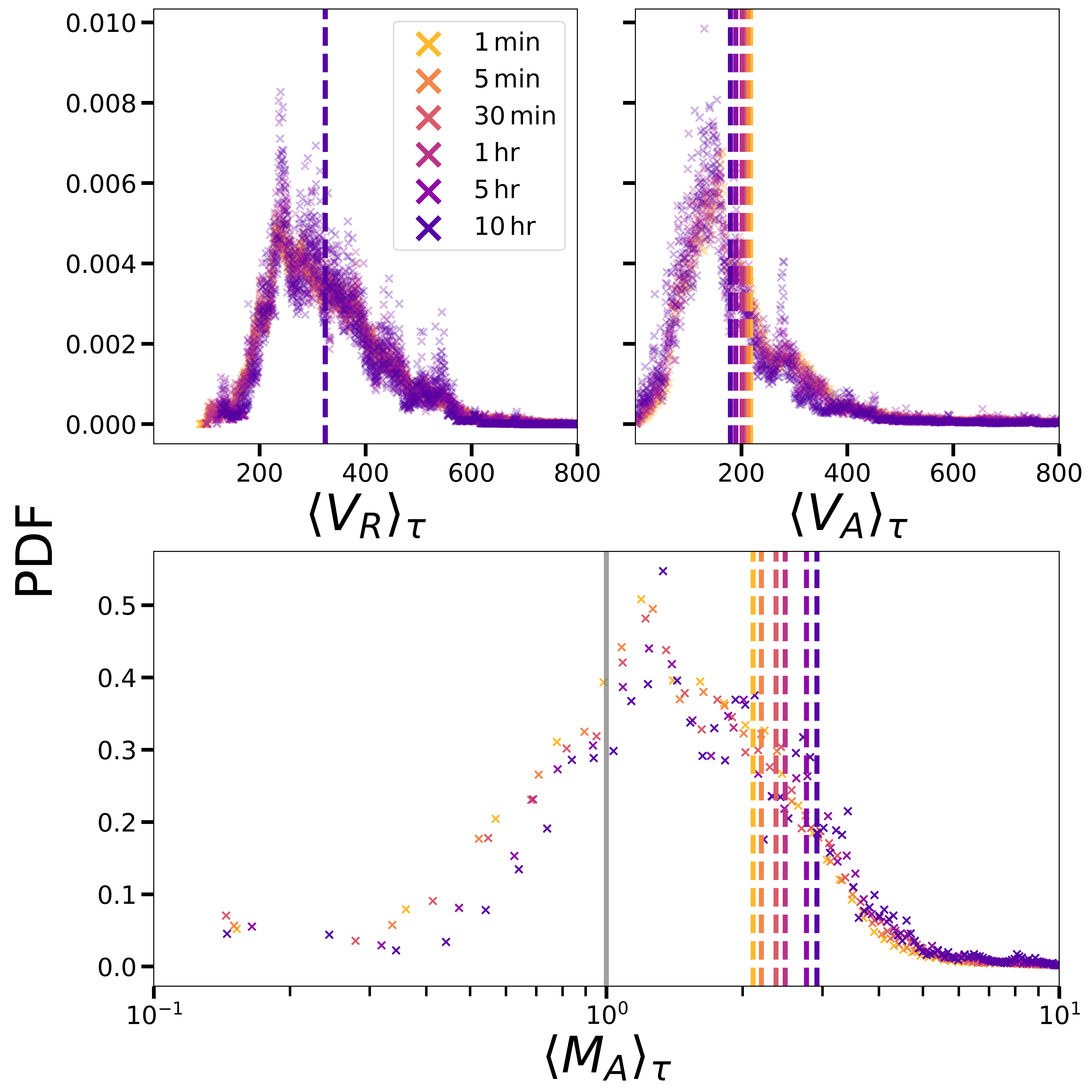}
\caption{
PDF of $\langle V_R \rangle_\tau$, $\langle V_A \rangle_\tau$, and $\langle M_A \rangle_\tau$ varying centered moving window mean of size $\tau$ for $E8-E19$. Dashed vertical lines indicate mean values. The solid vertical line indicates the Alfv\'enic transition. Width of rolling window average does not considerably change PDF or mean.
}
\label{fig:MA_pdf}
\end{figure}

\begin{figure}
\centering
\includegraphics[width=\linewidth]{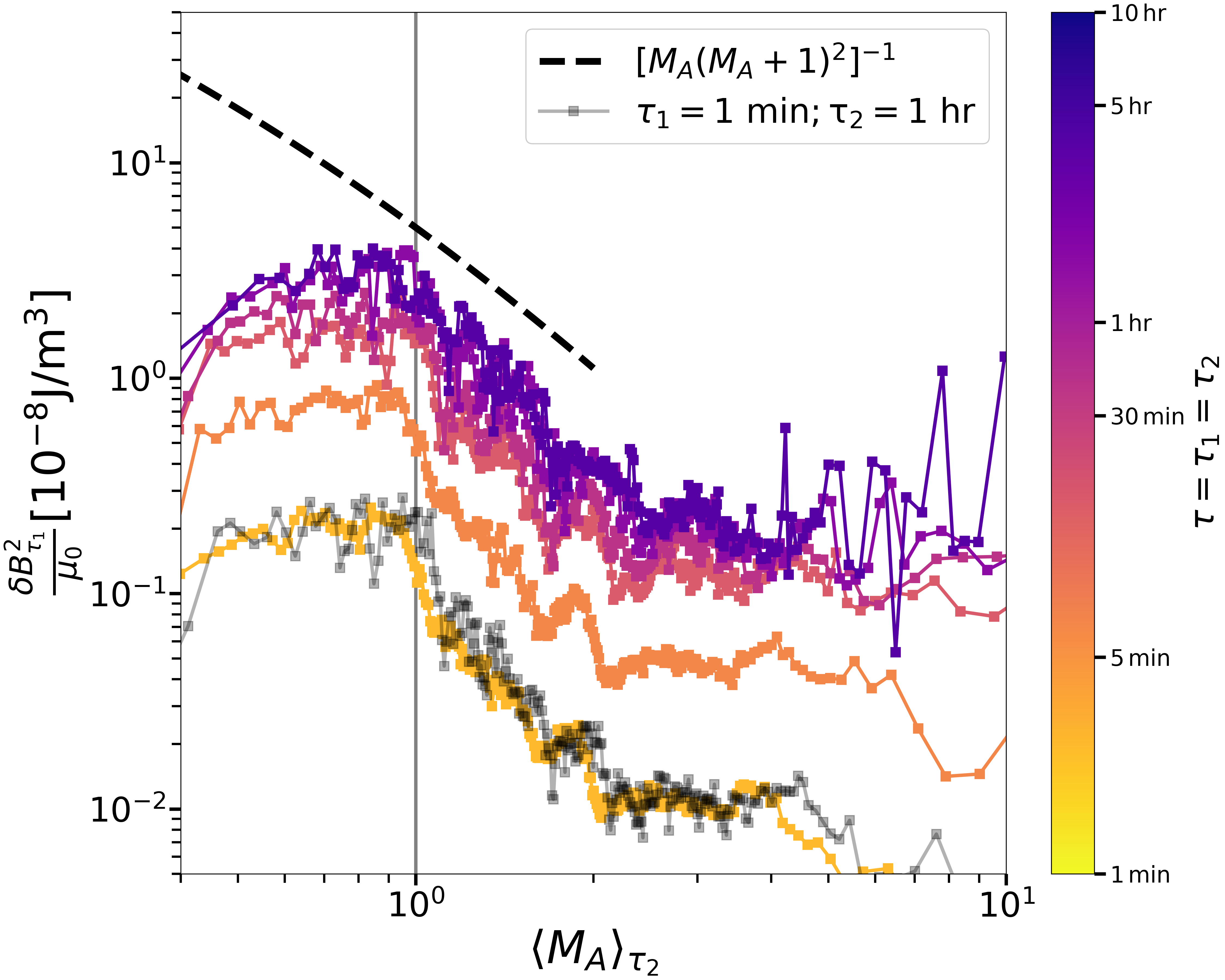}
\caption{
Binned median magnetic fluctuation energy density $\delta B^2_{\tau_1}/\mu_0$ vs $\langle M_A\rangle_{\tau_2}$ for various averaging timescales ($\tau_1, \tau_2$). All averaging windows where $\tau_1 = \tau_2$ are shown in the colorbar. All other averages are shown in the legend. The dashed line represents the trend expected from WKB theory for the case of non-interacting Alfv\'enic fluctuations generated at the Sun.
}
\label{fig:dB_MA_fit}
\end{figure}

\begin{figure}
\includegraphics[width=\linewidth]{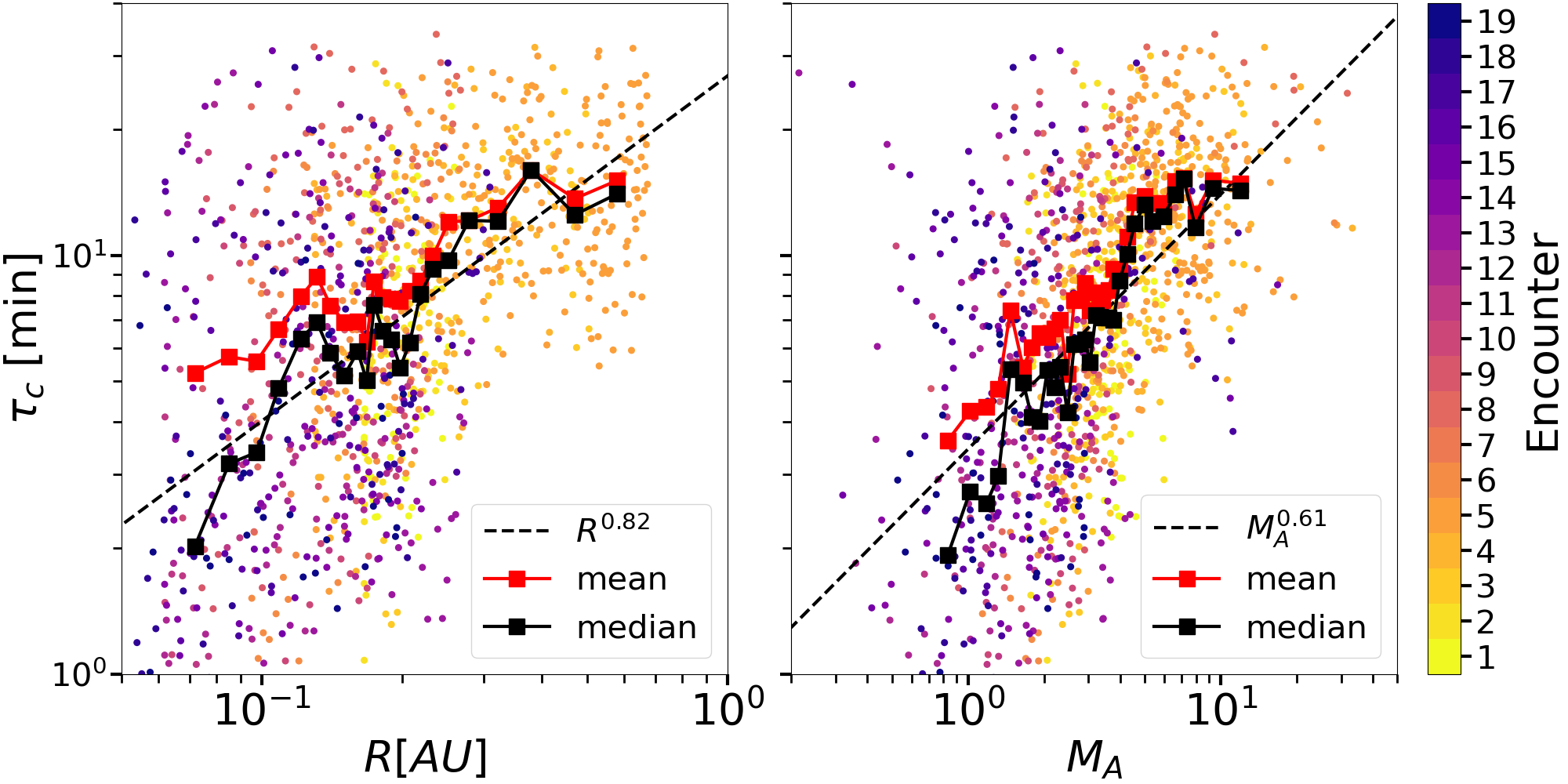}
\caption{Correlation times as a function of heliocentric distance (left) and \MA ~(right). The red and black squares are the mean and median values calculated over the binned intervals. Dashed lines are the best fits of the scatter.}
\label{fig:tau_c}
\end{figure}

Figure~\ref{fig:enc14} provides an overview of the time series of the (a) magnetic field components ($B_R, B_T, B_N$) (b) radial velocity ($V_R$) and electron density ($N_e$), (c) time in switchback state ($Z > 0.5$), (d) switchback parameter relative to $10\, \rm{hr}$ magnetic field ($Z_{10\,\rm{hr}}$) and (e)\MA\ $\langle M_A \rangle_\tau$ for PSP encounter 14. Panels (b) and (e) are calculated using rolling window of varying time intervals ($\tau = 1\, \rm{min},\,1\, \rm{hr}, \,5\, \rm{hr},\,\text{and}\, 10\,\rm{hr}$). The vertical lines in panel (c) display the intervals in switchback state where the color corresponds to the averaging length $Z_\tau$ the switchback interval is found. Encounter 14 is selected here as it has a long sub-Alfv\'enic duration. (Later, encounters 8-19 are used for Figures~\ref{fig:MA_pdf}-\ref{fig:dB_MA_fit}, while encounters 1-19 are used for Figure~\ref{fig:tau_c}.)
The horizontal dashed line in panel (e) represents the boundary separating the sub- and super-Alfv\'enic regimes.  Clearly, the choice of an averaging window interval does not considerably change the distribution of \MA. As a result, the duration of sub-Alfv\'enic regimes stays roughly the same. Furthermore, the fact that a majority of the intervals in a switchback state are in the super-Alfv\'enic regime, is almost independent of the size of averaging for both $M_A$ and $Z$.
For averaging intervals $\tau =$$1\, \rm{min}$, $1\, \rm{hr}$, $5\, \rm{hr}$, and $10\, \rm{hr}$, the total time the solar wind is in a switchback state for each of these is 0:00, 17:30, 32:20, and 47:40, respectively. When the reference ``background" field is too short to represent a
long duration average,
the switchback parameter does not properly measure local deflections of the field.

In Figure~\ref{fig:MA_pdf} we examine the probability distribution function (PDF) of the radial velocity $\langle V_R \rangle_\tau$, and Alfv\'en velocity $\langle V_A\rangle_\tau$ along with the \MA\ $\langle M_A \rangle_\tau$ calculated with various averaging intervals $\tau$. All these PDFs are calculated using PSP encounters 8-19 and displayed with equal binning. Each vertical dashed line represents the mean of the distribution corresponding to a 
specific averaging window $\tau$. Note that these means are almost unchanged with varying $\tau$. 
The mean of the \MA\ $M_A$ also changes only slightly, but is always super-Alfv\'enic i.e., $\langle M_A \rangle_\tau >1$. An important finding here is that the overall trend of these PDFs remain unchanged, suggesting that for a reasonable range of averaging intervals, the estimation of \MA\  is relatively independent of choice. Thus, the moderate changes in the averaging interval mainly affect the classification of individual intervals rather than the overall statistical properties of the solar wind.

Next, we investigate the variation of 
energy density with \MA, computed with 
varying definitions of the interval of averaging 
in Figure~\ref{fig:dB_MA_fit}. We display the variation of the (binned) 
magnetic fluctuation energy per unit volume ($\delta B_{\tau_1}^2/\mu_0)$ vs. $\langle M_A \rangle_{\tau_2}$ for PSP encounters 8-19, for various combinations of $\tau_1$ and $\tau_2$, which represent the averaging intervals for magnetic fluctuation energy and Alfv\'en Mach number, respectively.
In some cases we set $\tau_1=\tau_2$ to various values ranging from 1 min to 10 hr, and we also consider the case of $\tau_1=1$ min and $\tau_2=1$ hr. 
In all cases, the envelopes of the profiles are 
quite similar with a general trend toward higher energies with longer averages. This systematic increase in fluctuation energies for larger averaging intervals is expected as larger scale fluctuations are now included \citep{isaacs2015_Systematic}. 

Figure~\ref{fig:tau_c} shows the variation of the correlation time along the heliocentric distance (left) and \MA\ $M_A$ (right), extended for PSP encounters $1$ through $19$ to include additional super-Alfv\'enic intervals for comparison. Using the mean and the median values of the correlation times in the binned intervals, we find that $\tau_c$ increases moving farther from the sun. In addition, we find that the correlation time $\tau_c$ increases with $M_A$ suggesting a possible trend of correlation time with \MA.
We find that scaling with heliocentric distance follows an exponent of $R^{0.82}$, while the power-law scaling for \MA\ follows $M_A^{0.61}$, which are in qualitative agreement with previous findings \citep{ruiz2014_Characterization, cuesta2022_Intermittency}.

Note that we have not separated results according to the angle between the mean magnetic field and the radial direction; 
as found by \citet{cuesta2022_Isotropizationa} the 
PSP observations of correlation scale
close to the sun are systematically biased towards a preponderance of 
observations in the parallel direction.

\section{Conclusions and Discussion}\label{sec:discussion}
In this study, using PSP data selected from encounters $1$ through $19$, we investigate the effects of averaging intervals on \MA\ and turbulent properties such as energy densities, and correlation time. In particular, we find that, given the methodology as we have described above (e.g., Eq. \ref{eq:MA}), the detection of sub-Alfv\'enic intervals can be 
relatively insensitive to varying the length of the averaging interval, if the range of averaging is judiciously chosen based on the physics under consideration.
\begin{figure}
\includegraphics[width=\linewidth]{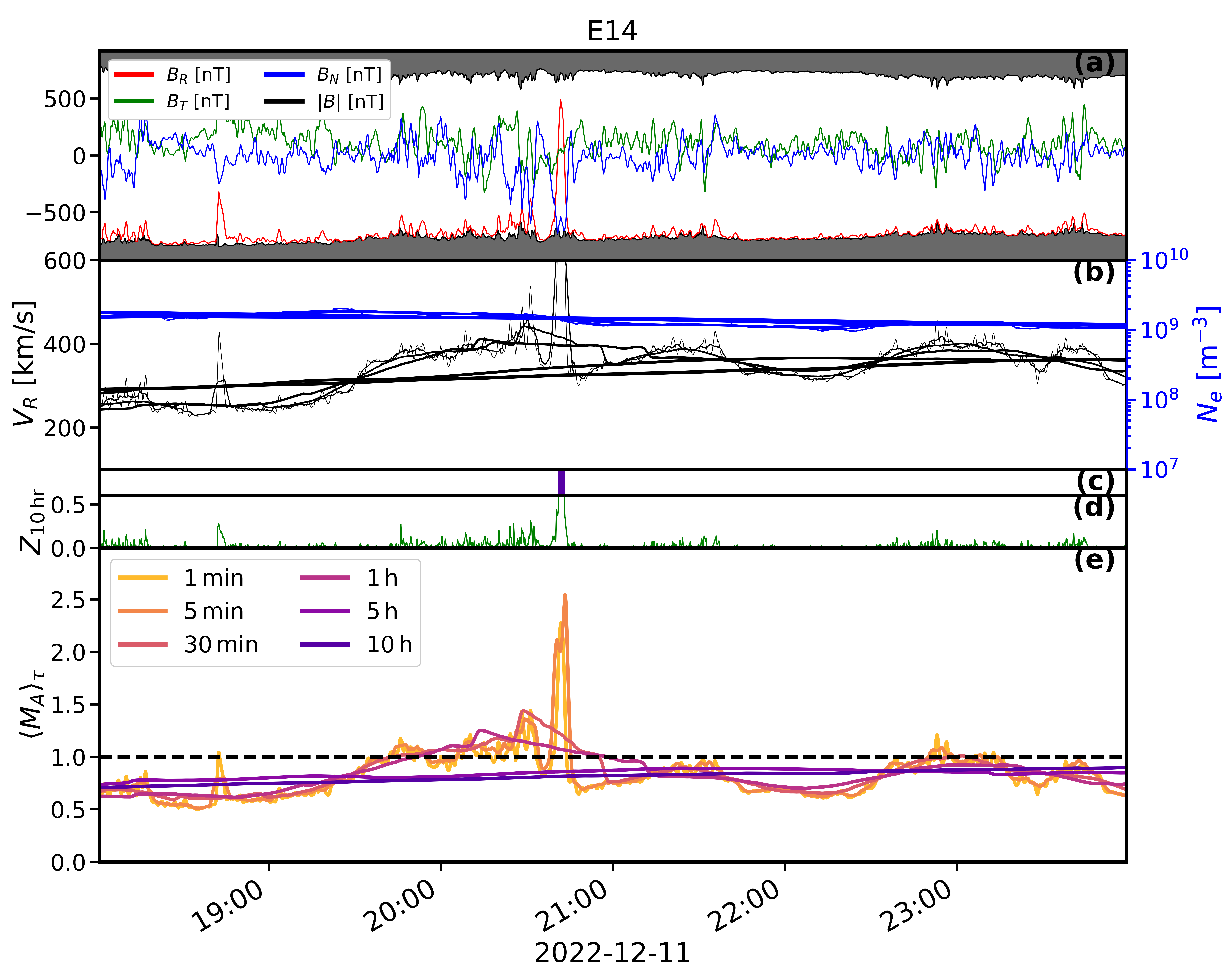}
\caption{
Zoom of Figure~\ref{fig:enc14} to highlight a brief excursion into 
super-Alfv\'enic conditions (e)
between 12-11 20:00
and 12-11 21:00. This period also includes a switchback (c) 
at approximately $\text{12-11 20:45}$. 
Note that use of 
large averaging windows ($5\,\rm{hr}$ or $10\,\rm{hr}$) 
suppress the identification of the 
local super-Alfv\'enic state seen in all the smaller averaging windows
($1\,\rm{m in}$ to $1\,\rm{hr}$).
}
\label{fig:MA_enc_zoom}
\end{figure}

Notably, we find that the 
distributions of $M_A$ computed from encounters 8-19 
are essentially
invariant to varying the averaging between $1\,\rm{min}$ and $10\,\rm{hr}$. 
Most of the switchbacks 
observed, for example in Figure~\ref{fig:enc14}, are found in super-Alfv\'enic flows, and for almost all of them, varying the averaging interval does not change their status as super-Alfv\'enic. 
This is confirmed in diagnostics for all encounters that we analyzed, by inspection of plots constructed in the same format 
as Figure~\ref{fig:enc14}. (These are not shown here.) 
In this way, we confirm
what is widely accepted at this point - namely that 
switchback intervals remain predominantly a component of the super-Alfv\'enic wind~\citep{pecora2022_Magnetic, bandyopadhyay2022_SubAlfvenica, jagarlamudi2023_Occurrence, payne2026_Evolution, goodwill2026_Parker}.

On the other
hand, individual switchback identifications events can be influenced by choice of averaging interval. It has been
noted~\citep{gosling2009_ONESIDED,gosling2011_PULSED,kasper2019_Alfvenic, matteini2014_Dependence, sioulas2025_Propagation, badman2026_Properties, sioulas2026_Generation} 
that enhancements of solar wind velocity
are often associated with 
sharp reversals in magnetic field, 
corresponding to a local 
increase in \MA. 
Such fluctuations actually form the basis of the spatially complex ``Alfv\'en zone'' 
suggested by \citet{chhiber2022_extended}.
Switchbacks generally appear in super-Alfv\'enic wind, 
where the Chandrasekhar criterion for Kelvin-Helmholtz rollup can be more easily achieved~\citep{ruffolo2020_Sheardriven}.
Given that 
shear instabilities are inherently local, 
it seems 
natural to expect the possibility that some switchbacks could
form in regions that are 
temporarily super-Alfv\'enic due to fluctuations as argued by \citet{chhiber2022_extended} and \citet{cranmer2023_Suns}.

Interestingly, 
some recent works argue in a completely contrary way: 
In the review by 
\citet{badman2026_Properties}, the same interval we consider in Figure~\ref{fig:enc14} is examined and it is argued that the temporary excursions into the super-Alfv\'enic state can (and perhaps, should) 
be eliminated by computing a \MA\ using longer averaging
intervals. 
In contrast to the global statistics which are fairly stable to changes in 
averaging (see Figure \ref{fig:MA_pdf}), it seems clear that brief excursions in super-Alfv\'enic flow (or even the reverse) can be eliminated by longer averaging. 
This is illustrated in 
Figure~\ref{fig:MA_enc_zoom} 
that examines a subset of the period in Figure~\ref{fig:enc14}. This is also a period discussed in Fig. 20 of \citet{badman2026_Properties}. 
As we show in this figure, 
an excursion into a super-Alfv\'enic state, along with a switchback, occurs near 12-11 20:45.
The temporarily super-Alfv\'enicity remains evident even in the 30 minute and 1 hour definitions. 
However, using the 5 hour and 10 hour definitions, the  
curve remains sub-Alfv\'enic throughout. Now we may recall (see Figure~\ref{fig:tau_c}) that the correlation time for such a period is around $4\,\rm{min}$ to $5\,\rm{min}$. 
Associating this with the correlation scale, it seems appropriate to postulate that shear effects would be most evident at scales of a few minutes, and the local dynamics would be less sensitive to characteristics obtained by averaging over $5$ to $10$ hours, which is many times the inferred correlation scales.  
Therefore we conclude that the excursion into super-Alfv\'enic flow near
12-11 20:45 is physically relevant. 

We note in passing that near 12-11 23:00 in Figure ~\ref{fig:MA_enc_zoom}, there is another smaller excursion into a super-Alfv\'enic state, again visible in $1\,\rm{min}$ and $5\,\rm{min}$ averaging. It is again eliminated in the 5 hour and 10 hours definitions of \MA. There is no switchback in this period, nor is there any implication there that there should be. 
For example, there are other conditions to form a switchback according to the developments in~\citet{ruffolo2020_Sheardriven}, including the 
Chandrasekhar criterion.

To further emphasize the 
necessity of a judicious choice of averaging intervals, let us examine the effects of increasing the averaging window even further.
If we were to compute $M_A$ by 
averaging over several days, 
it is clear (see Figure~\ref{fig:enc14}) 
that sub-Alfv\'enic intervals would no longer be present.
Whatever local 
physics may be attributed to those sub-Alfv\'enic conditions
would be missed. 
Incorrectly chosen averaging 
can obviously obscure the 
physics. 
For many purposes, reference to the correlation scale may provide guidance. For other purposes such as characterizing a macroscopic feature such as a solar wind stream, or the average Parker spiral field, much longer averaging (even up to a solar rotation period) may be indicated. These results show that there is no universally optimal averaging interval. Rather, the appropriate averaging scale depends on the physical process being investigated, with shorter intervals better capturing local plasma dynamics and longer intervals providing a more representative description of the large-scale solar wind.

Analogous statements may be made with 
regard to the behavior of magnetic energy density. 
While the amplitude of the magnetic energy density increases with a larger averaging intervals \citep{isaacs2015_Systematic}, the overall trend of the fluctuation energy density (Figure~\ref{fig:dB_MA_fit})
remains consistent with variation of the 
averaging interval from $1\,\rm{min}$ to $10\,\rm{hr}$. 
In particular, the monotonic decrease of the fluctuation energy density with decreasing $M_A$ in the super-Alfv\'enic region is seen for all cases. 
In addition, in the sub-Alfv\'enic regime, 
departures from 
predicted WKB scalings \citep{jacques1978_Solar, ruffolo2024_Observed} 
are seen in all cases when
varying the
averaging interval from $1\,\rm{min}$ to $10\,\rm{hr}$. This contradicts the idea that fluctuations are predominantly generated at the Sun and travel outward; the rise in the measured fluctuation energy relative to the WKB trend proves that most of the super-Alfv\'enic fluctuation energy is generated in the range $0.5 < M_A < 1$.
If one were to further increase the averaging intervals to periods 
as long as several days or weeks, these
features would be likely obscured, because such long averaging intervals could significantly reduce the time for which PSP is imputed to be within sub-Alfv\'enic solar wind.

The results of the present paper are intended to 
provide insight into 
the rationale for 
selecting various averaging windows in defining a parameter such as Alfv\'en Mach number. 
As such, the definition of a ``background'' may depend on the associated problem at hand,
such as the largest scale structure that contributes to the dynamics under consideration. 
For the purpose of the 
local turbulence
processes in the solar wind, the background may 
often be associated with the correlation time or length. This background definition is particularly key when referencing fluctuations to identify features within the turbulent dynamics, such as switchbacks, shocks, and solar wind expansion.

\section*{Availability Statement}
Data used in figures is published at \href{}{https://doi.org/10.5281/zenodo.21340319}, which follows methods outlined by \citet{Moncuquet_2020, romeo2023_NearSun, ruffolo2024_Observed}. Scripts used for analysis and producing figures are available at \newline (\href{}{https://github.com/jgoodwill1/Goodwill2026GRL}).

\section*{Acknowledgements}
This work was supported at the University of Delaware in part by the PSP/IS\(\odot\)IS project through subcontract SUB0000165 and from Princeton and PUNCH project through subcontract N99054DS from NASA/SWRI to the University of Delaware. This work was also partially 
supported by the U.S. National Science Foundation, award PHY-2108834, through the NSF/DOE Partnership in Basic Plasma Science and Engineering.
J.~G.\ was supported by the Delaware NASA Space Grant program grant number 80NSSC20M0045 at the University of Delaware. 
M.~E.~C.\ was supported by the NASA LWS grant NNN06AA01C for PSP/IS\(\odot\)IS. 
D.~R. and P.~P.\ were supported from the NSRF via the Research and Innovation Acceleration Agency for Competitiveness and Area Development (RCAD) (Program Management Unit for Frontier Brainpower and Future Industries) [grant number B39G690003].
R.~C. was supported by the NASA  Living With a Star (LWS) Science program grant 80NSSC22K1020.


\end{document}